\documentclass[aps,showpacs,preprint]{revtex4}%
\usepackage{amsfonts}
\usepackage{amsmath}
\usepackage{amssymb}
\usepackage{graphicx}%
\begin{document}
\preprint{PZJ/2009.July}
\title[Short title for running header]{Forced Granular Orifice Flow}
\author{Zheng Peng}
\affiliation{School of Physics, Central South University, Changsha 410083, China}
\author{Hepeng Zheng}
\affiliation{School of Physics, Central South University, Changsha 410083, China}
\author{Yimin Jiang}
\affiliation{School of Physics and State Key laboratory for Powder Metallurgy, Central
South University, Changsha 410083, China}
\begin{abstract}
The flow of granular material through an orifice is
studied experimentally as a function of force $F$
pushing the flow. It is found that the flow rate
increases linearly with $F$ --- a new, unexpected
result that is in contrast to the usual view that $F$,
completely screened by an arch formed around the
orifice, has no way of altering the rate. Employing
energy balance, we show that this behavior results
mainly from dissipation in the granular material.

\end{abstract}
\volumeyear{year}
\volumenumber{number}
\issuenumber{number}
\eid{identifier}
\date[Date text: ]{\today}
\startpage{1}
\endpage{2}

\maketitle

Although granular materials such as sand, rice etc.
are familiar in daily life and industrial handling,
their static and dynamic behavior is only beginning to
be understood\cite{G-Phys}. In these systems, arch
formation is an important effect, which may remarkbly
influence the mechanical properties of these
materials\cite{Duran}. A classic example of a dynamic
arch can be found in the simple system of a silo
(fig.1a), in which mass is discharged from an orifice.
Since ancient times, it has been supposed (and used
for making "hour glasses") that the discharge rate $Q$
of the granular orifice flow (GOF) is independent of
the height $H$ of the silo column --- a phenomenon
commonly believed to be due to the so called "free
fall arch" (FFA) formed around the
orifice\cite{Nedderman}. Moreover, if the orifice
diameter $D$ decreases, the FFA may jam into an static
arch, at which point the flow stops. The jamming
transition was studied in \cite{To} for the 2D case.
Specifically the FFA is described as a semispherical
surface spanning the orifice, the total energy of
which is taken as minimal in the "minimum energy
theory" \cite{FFA}, and the normal stress taken as
vanishing in the "hour glass theory" \cite{HGT}.

These theories represent the most advanced continuous
modeling of GOF, from which the empirical Beverloo
formula for $Q$ can be derived\cite{Nedderman}. The
FFA assumed here is rather unusual. It can resist all
force $F$ given by the weight of the column $H$, and
any load $L$ on its top, such that $Q$ is fully
protected from being influenced. Clearly, this basic
assumption can be directly checked, by measuring the
rate $Q$ as a function of $F$. We performed this
measurements and found a linear
dependence (if $F$ is not too small),%
\begin{equation}
Q=A\left(  1+\alpha F\right)  \label{1}%
\end{equation}
instead $\partial Q/\partial F=0$. This demonstrates
that the properties claimed for FFA in\cite{FFA,HGT}
are, at the least, not rigorously valid. The
difference between (\ref{1}) and the Torricelli's law
for ideal liquids: $Q\sim\sqrt{F}$ is clearly a result
of granular arch and dissipation. Note that to measure
the variation of $Q$ with $F$, we need to remove the
Janssen effect\cite{Hagen,Janssen}, that the load $L$
and the weight of  sand are typically redirected onto
the side wall of silo, steering $F$ to a saturation
value $F_{sat}$ determined by the frictional property
between sand and wall, including the history of the
preparation. The screening of $F$ can be removed, for
instance, by allowing the side wall to move freely and
vertically (see fig.1b), then $F$ is a sum of the load
$L$, the weight of sand $M$, and the wall $M_{W}$ -- a
quantity that can be experimentally controlled and
varied.

%

%TCIMACRO{\FRAME{ftbFU}{2.7017in}{2.5529in}{0pt}{\Qcb{Granular flow from a
%bottom orifice with (a): fixed wall, and (b): vertically movable wall. (See
%text).}}{}{f1.eps}{\special{ language "Scientific Word";  type "GRAPHIC";
%width 2.7017in;  height 2.5529in;  depth 0pt;  original-width 6.3157in;
%original-height 5.9655in;  cropleft "0";  croptop "1";  cropright "1";
%cropbottom "0";  filename 'f1.EPS';file-properties "XNPEU";}}}%
%BeginExpansion
\begin{figure}
[b]
\begin{center}
\includegraphics[
height=2.5529in,
width=2.7017in
]%
{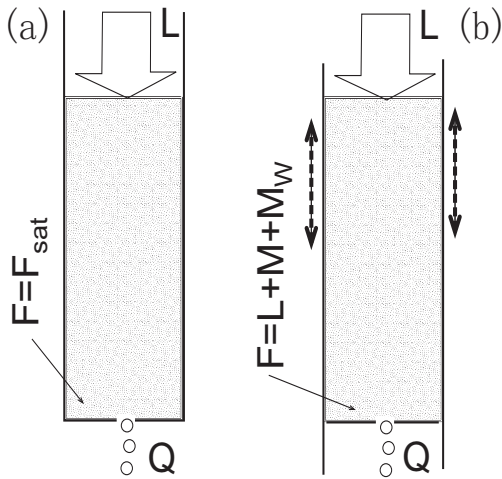}%
\caption{Granular flow from a bottom orifice with (a): fixed wall, and (b):
vertically movable wall. (See text).}%
\end{center}
\end{figure}
%EndExpansion

The variation of $Q$ with the orifice diameter $D$ has
been studied for a long time~\cite{Beverloo,Hagen}.
The Beverloo formula,
\begin{equation}
Q=C\rho\sqrt{g}\left(  D-D_{a}\right)  ^{5/2} \label{2}%
\end{equation}
though well known, needs as a topic of current
interests, validation and a clarification of its
limitation, see e.g.~\cite{Mankoc}. Here, $g$ is the
gravitational acceleration, $C$ a parameter, and
$D_{a}$ a correction to the orifice diameter of an
empty annulus, necessary because the flow has a
tendency to avoid the periphery of the
orifice~\cite{FFA,Nedderman}. More recently, Sheldon
and Durian measured the variation of $Q$ with the
orifice tilt angle $\theta$~\cite{Durian}. Neither
dependence (not to mention their combination) is well
understood, posing a serious challenge to any granular
theory. We shall consider only horizontal orifice
planes ($\theta=0$), focusing our attention on the
variation of $Q$ with the pushing force $F$. To our
best knowledge, this property has not been addressed
before. There is a simulation which shows an
increasing of the GOF velocity with column
height~\cite{Hirshfeld}, which seems similar to
(\ref{1}), but its authors attributed this effect to
the choice made in the simulation of the damping
parameter, and did not take it as a general behavior
of GOF. Moreover, strong vibration may break up the
arches, dramatically altering the flow
behavior, even rendering it to become the Torricelli
one~\cite{vibration,Martinez}.%
Also, for dilute granular flows studied in\cite{Hou},
a deviation from (\ref{1}) is expected, because no
arches are formed.

Experimental setup is shown in fig.1b, where the side wall is a PVC cylinder
of inner diameter $\phi=10.4$ cm, weight $M_{W}=1.16$ kg. It is horizontally
restricted by pulley wheels such that it can be moved freely in the vertical
direction. The granular material consists of nearly mono-disperse glass beads
of the diameter $d\sim3$ mm. Bottom orifice is of $D=14$, $16$, $18$, $20$,
$22$ mm. Load $L$ ranges from $0$ to about $40$ kg. For each experiment, the
flow starts with a granular column of the height $H_{i}=51$ cm, prepared by
direct pouring, and ends with $H_{f}=37$ cm, the total flowed mass is about
$M_{0}=1.8$ kg. The bulk density $\rho$ of the granular sample is about $1.5$
g/cm$^{3}$. GOF rate $Q$ is measured by continuing recording (15 records/s)
the flowed mass $M=\int_{0}^{t}Qdt^{\prime}$ with an electronic balance of the
precision $\pm0.5$ g. It is observed that $M\left(  t\right)  $ is a fairly
good straight line, indicating a constant rate $Q$ given by its slope (see
e.g. fig.2 inset). For every measurement the flow rate is found reproducible,
with a fluctuation less than 1\%. First, we performed measurements with a
fixed cylinder without load ($L=0$), and obtained a rate dependence
$Q_{\text{fixed wall}}^{L=0}(D)$ that follows the Beverloo law~(\ref{2})
excellently, as shown in fig.2. Best fitting gives $D_{a}\simeq3.7$ mm and
$C\simeq0.51$, close to the values $4.5$ and $0.58$ given in the
book~\cite{Nedderman}.%

%TCIMACRO{\FRAME{ftbpFU}{3.3399in}{2.2157in}{0pt}{\Qcb{$Q^{2/5}$ versus $D$.
%Full straight is the best linear fit. Inset shows a measured discharging
%$M(t)$.}}{}{f2.eps}{\special{ language "Scientific Word";  type "GRAPHIC";
%width 3.3399in;  height 2.2157in;  depth 0pt;  original-width 3.4039in;
%original-height 2.4111in;  cropleft "0";  croptop "1";  cropright "1";
%cropbottom "0";  filename 'f2.EPS';file-properties "XNPEU";}}}%
%BeginExpansion
\begin{figure}
[ptb]
\begin{center}
\includegraphics[
height=2.2157in,
width=3.3399in
]%
{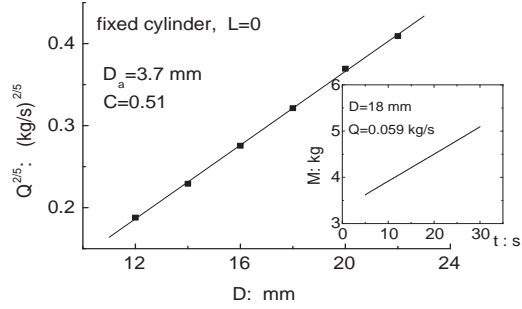}%
\caption{$Q^{2/5}$ versus $D$. Full straight is the best linear fit. Inset
shows a measured discharging $M(t)$.}%
\end{center}
\end{figure}
%EndExpansion

Next, we release the cylinder, allowing it to move vertically, such that the
silo screening effect is eliminated. In this situation, the bottom force $F$
pushing the flow is given by the total mass of the setup, which varies
slightly with the time, $M=M\left(  t\right)  $. During the flow, which starts
at $t=0$ and terminates at $t_{f}$, the temporally averaged pushing force is%
\begin{equation}
\left\langle F\right\rangle =\frac{1}{t_{f}}\int_{0}^{t_{f}}Fdt=L+M_{W}%
+\frac{\left(  H_{i}+H_{f}\right)  }{2\left(  H_{i}-H_{f}\right)  }%
M_{0}\text{.} \label{4}%
\end{equation}
Because no noticeable influences of this slight temporal dependence of $F$ on
the rate $Q$ has been observed, (the measured $\int Qdt$ is always a good
straight line as given in fig.2 inset), we shall assume $F\approx\left\langle
F\right\rangle $ in what follows. Fig.3 shows the measured variation of $Q$
with $\left\langle F\right\rangle $, all straight lines, confirming
Eq.(\ref{1}) with $\alpha>0$. Note that although the load, the cylinder and
the upper part of the granular column go down slowly during the flow as a
whole, no relative motion among them is observed.%

%TCIMACRO{\FRAME{ftbpFU}{3.339in}{2.5815in}{0pt}{\Qcb{Measured GOF rate $Q$
%versus average pushing force $\left\langle F\right\rangle $ (symbols), and
%their best linear fit (straights), for various orifice diameters $D$.}}%
%{}{f3.eps}{\special{ language "Scientific Word";  type "GRAPHIC";
%width 3.339in;  height 2.5815in;  depth 0pt;  original-width 4.0534in;
%original-height 2.7017in;  cropleft "0";  croptop "1";  cropright "1";
%cropbottom "0";  filename 'f3.eps';file-properties "XNPEU";}}}%
%BeginExpansion
\begin{figure}
[ptb]
\begin{center}
\includegraphics[
height=2.5815in,
width=3.339in
]%
{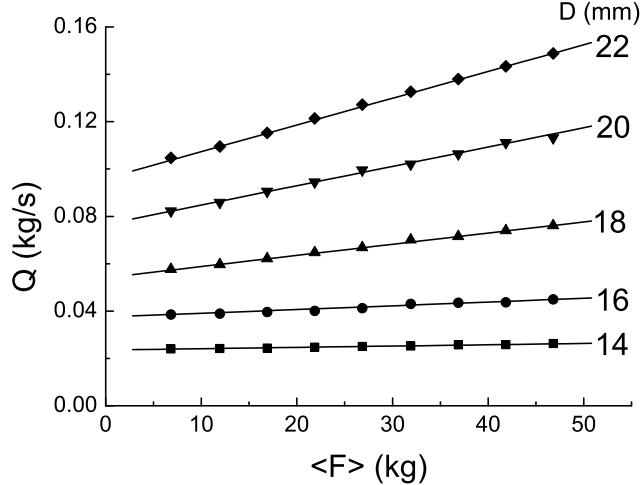}%
\caption{Measured GOF rate $Q$ versus average pushing force $\left\langle
F\right\rangle $ (symbols), and their best linear fit (straights), for various
orifice diameters $D$.}%
\end{center}
\end{figure}
%EndExpansion

To check the validity of the Beverloo law (\ref{2}) for pushed GOF, we plot
$Q^{2/5}$ versus $D$ for various forces $\left\langle F\right\rangle $ in
fig.4a. Again linearity was obtained, implying a good agreement. By best
linear fit, we found that both Beverloo parameters $C$, $D_{a}$ increase with
pushing force (fig.4b and c), with the former showing a linear behavior,
\begin{equation}
C=C_{0}\left(  1+\left\langle F\right\rangle /F_{0}\right)  \label{5}%
\end{equation}
(where $C_{0}=0.44$ and $F_{0}=35$ kg), while the latter is nonlinear,
displaying a saturation behavior as $\left\langle F\right\rangle $ increases,
approximately as
\begin{equation}
D_{a}=D_{\infty}-\frac{D_{\infty}-D_{0}}{1+\left\langle F\right\rangle
/F_{0}^{\ast}} \label{6}%
\end{equation}
(where $D_{\infty}$ is its biggest value at the large force limit
$\left\langle F\right\rangle \rightarrow\infty$, and $D_{0}$ the smallest one
at the zero force limit). Best fit yields $F_{0}^{\ast}\simeq F_{0}$,
$D_{\infty}\simeq8.3$ mm, $D_{0}\simeq3.2$ mm, see fig.4c.

In the Beverloo law, $D_{a}$ accounts for clogging, which happens when the
orifice is mesoscopic, of a size no more than a few grains. Our measurements
show that clogging is slightly more likely with increased force. Note that as
a finite size effect clogging cannot exist in the macroscopic limit, $D>>d$,
where the Coulomb yield law holds. This means $D_{a}$ must saturate at a
mesoscopic size.

%

%TCIMACRO{\FRAME{ftbpFU}{3.3399in}{2.5131in}{0pt}{\Qcb{(a): $Q^{2/5}$ versus
%$D$ for various pushing force $\left\langle F\right\rangle $. Symbols are the
%data in fig.3, and full straights their best linear fits. Their Beverloo
%parameters are plotted in (b) and (c) (symbols) as functions of $\left\langle
%F\right\rangle $, where full curves are the model (\ref{5},\ref{6}).}}%
%{}{f4.eps}{\special{ language "Scientific Word";  type "GRAPHIC";
%width 3.3399in;  height 2.5131in;  depth 0pt;  original-width 3.3762in;
%original-height 2.4967in;  cropleft "0";  croptop "1";  cropright "1";
%cropbottom "0";  filename 'f4.eps';file-properties "XNPEU";}}}%
%BeginExpansion
\begin{figure}
[ptb]
\begin{center}
\includegraphics[
height=2.5131in,
width=3.3399in
]%
{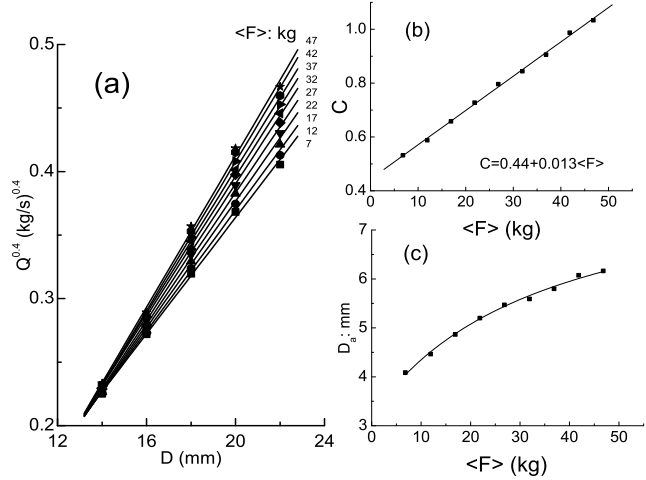}%
\caption{(a): $Q^{2/5}$ versus $D$ for various pushing force $\left\langle
F\right\rangle $. Symbols are the data in fig.3, and full straights their best
linear fits. Their Beverloo parameters are plotted in (b) and (c) (symbols) as
functions of $\left\langle F\right\rangle $, where full curves are the model
(\ref{5},\ref{6}).}%
\end{center}
\end{figure}
%EndExpansion

In the above generalization of the Beverloo law which includes the pushing
force, variations of $C$, $D_{a}$ as modeled by (\ref{5},\ref{6}) is
considered. Unfortunately, this way of generalization obscures the linear
relation between $Q$ and $F$. Alternatively, we may consider a different
generalization that keeps the linearity: Letting $A$ and $\alpha\ $ in
(\ref{1}) vary with $D$ as
\begin{align}
A  &  =\rho\sqrt{g}C_{0}\left(  D-D_{0}\right)  ^{5/2},\label{7}\\
\alpha &  =\left(  1-\frac{D_{1}}{D}\right)  \frac{1}{F_{0}}. \label{8}%
\end{align}
When the model constants $C_{0}$, $D_{0}$, $F_{0}$ take the same values as in
(\ref{5},\ref{6}), the two models become identical in the limit $\left\langle
F\right\rangle \rightarrow0$ or $D>>D_{0,\infty,1}$. As the difference between
them is very small (see fig.5a), it can not be decided with the present data
which version of generalization is more appropriate. Using the values for $A$,
$\alpha$ given by the straight lines of fig.3, we obtain a good agreement with
(\ref{7},\ref{8}), if $D_{1}\simeq13$ mm (fig.5b,c).

It is worth noting that the above empirical formulas for $Q\left(  F,D\right)
$ are invalid in the limits of small pushing forces $F\rightarrow0$ and
clogging orifice diameter $D\rightarrow D_{\text{clog}}>D_{\infty}$, where $Q$
is vanishing but the formulas do not show it. The diameter $D_{\text{clog}}$,
at which GOF is clogged, is $F$ dependent, and not a parameter of these
models. For the experiments of this work, we have $D_{1} >D_{\text{clog}%
}>D_{\infty}>D_{0}$. That $D_{\text{clog}}$ could not be obtained
with the models is probably due to the fact that the transition from
flow to clogging is not a sharp one, but occurs via an intermittent
flow. The above values for $D_{0,\infty}$ is well below the
transition at which no flow is possible, while $D_{1}$ belongs to
the intermittent regime, which starts from $D\sim12$ mm for $L=0$,
and $\sim14$ mm for $L=40$ kg. Moreover, in addition to causing the
flow, $F$ also increases clogging, as it jams the grains around the
orifice and FFA. At $D\sim D_{1}$, these two effects compensate each
other, and we have $\alpha\sim0$, implying $Q$'s independence of
$F$.

%

%TCIMACRO{\FRAME{ftbpFU}{3.3399in}{2.7146in}{0pt}{\Qcb{(a): Comparison between
%the model (\ref{5},\ref{6}) (circls) and the model (\ref{7},\ref{8}) (curve).
%Inset is an ampified figure for small $D$. (b): variation of $\alpha$ , (c):
%of $A$, with $D$. Symbols are measurements, and curves are the model
%(\ref{7},\ref{8}).}}{}{f5.eps}{\special{ language "Scientific Word";
%type "GRAPHIC";  maintain-aspect-ratio TRUE;  display "USEDEF";
%valid_file "F";  width 3.3399in;  height 2.7146in;  depth 0pt;
%original-width 4.3586in;  original-height 2.6984in;  cropleft "0";
%filename 'f5.eps';file-properties "XNPEU";}}}%
%BeginExpansion
\begin{figure}
[ptb]
\begin{center}
\includegraphics[
height=2.7146in,
width=3.3399in
]%
{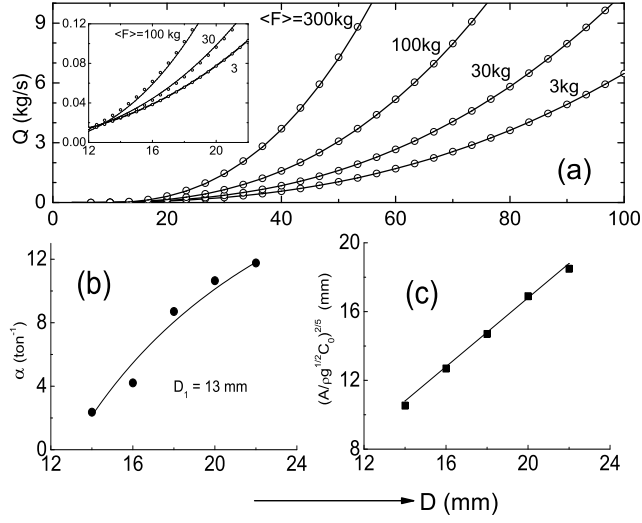}%
\caption{(a): Comparison between the model (\ref{5},\ref{6}) (circls) and the
model (\ref{7},\ref{8}) (curve). Inset is an ampified figure for small $D$.
(b): variation of $\alpha$ , (c): of $A$, with $D$. Symbols are measurements,
and curves are the model (\ref{7},\ref{8}).}%
\end{center}
\end{figure}
%EndExpansion

The dependence of the GOF rate on the pushing force $F$ can be qualitatively
understood employing the energy balance: The power $W_{in}$ injected into the
granular system is equal to the sum of the power $W_{out}$ carried by the flow
and the heat production $W_{D}$ inside it:%
\begin{equation}
W_{in}=W_{D}+W_{out}\text{.} \label{9}%
\end{equation}
If the height of the granular column is not too small, we have,
\begin{equation}
W_{in}=\frac{4gFQ}{\pi\rho\phi^{2}}\text{, }W_{out}=\frac{8Q^{3}}{\pi^{2}%
D^{4}\rho_{out}^{2}}+W_{out}^{fluct.}. \label{11}%
\end{equation}
The contribution $W_{out}^{fluct.}$ accounts for the kinetic energy
of the fluctuating motion of the grains, which we assume is much
smaller than the first contribution and may hence be ignored;
$\rho_{out} (<\rho)$ is the density of the orifice flow. For ideal
liquid, $W_{D}=0$, and the balance is given between $W_{in}$ and
$W_{out}$, resulting in the square root dependence $Q\sim\sqrt{F}$
of Torricelli. But for GOF, dissipation is so strong that the
balance is mainly between $W_{in}$ and $W_{D}$. With the present
experiment, (assuming $\rho_{out}=\rho/2$,) we estimate
$W_{out}/W_{in}\lesssim10^{-2}$, indicating that $W_{out}$ is
negligible. In this situation, the rate $Q\left(  F\right) $ is
determined by the dependence of the heat production power on the
pushing force, $W_{D}\left(  F\right)  $. Tailor expanding and
comparing the result
with (\ref{1}), we have%
\begin{equation}
W_{in}=W_{D}=W_{D}^{(1)}F+W_{D}^{(2)}F^{2}+... \label{12}%
\end{equation}
and%
\begin{align}
A  &  =\pi\phi^{2}\rho W_{D}^{(1)}/4g\text{, }\label{13}\\
\alpha &  =W_{D}^{(2)}/W_{D}^{(1)}\text{.} \label{13a}%
\end{align}
Further study needs to consider the heat produced in the system. Employing
Granular Solid Hydrodynamics (GSH) as given in \cite{GSH}, we show a
order-of-magnitude estimate for $W_{D}^{(1)}$ below.

First, note that in the present setup, dissipation occurs only in a small
volume $V_{D}$ close to the orifice at the bottom. Assuming that this region
is a dense granular gas with a granular temperature $T_{g}$, we have,
according to GSH, the following relations for the pressure and heat production
(up to quadratic order in $T_{g}$)
\begin{align}
P  &  =\frac{a\rho}{2\rho_{c}}\frac{\rho b_{0}T_{g}^{2}}{\left(  1-\rho
/\rho_{c}\right)  ^{1-a}}\label{14}\\
W_{D}  &  =V_{D}\gamma_{0}T_{g}^{2} \label{15}%
\end{align}
where $\rho_{c}\sim1.0667\rho$ is the random closest packing density, the
exponent $a\sim0.1$, and $b_{0},\gamma_{0}$ material parameters, of granular
matter. Eliminating $T_{g}$ with Eqs.(\ref{14},\ref{15}) and noting that
$P\sim4gF/\pi\phi^{2}$ we get%
\begin{equation}
W_{D}=V_{D}\gamma_{g0}\frac{8\left(  \rho_{c}-\rho\right)  g}{a\pi\phi^{2}%
\rho}F \label{16}%
\end{equation}
where%
\begin{equation}
\gamma_{g0}=\frac{\gamma_{0}}{\rho b_{0}\left(  1-\rho/\rho_{c}\right)  ^{a}}
\label{17}%
\end{equation}
can be considered as a constant of about $0.5$ Hz  \cite{GSH}.
Inserting $W_{D}^{(1)}$ as
given by (\ref{16}) into (\ref{13}) we get%
\begin{equation}
A=2\gamma_{g0}V_{D}\left(  \rho_{c}-\rho\right)  /a. \label{18}%
\end{equation}
This $A$ is comparable to the measurements if we take $V_{D}$ as $10$ times
the orifice volume $D^{3}$. To estimate $\alpha$, higher order expansion terms
are needed. The detailed analysis will be presented separately.

Orifice flows are generally pushed by the force
difference between the in- and outside of the orifice.
The increase of the flow rate with the force obeys the
Torricelli's law if an arch is absent, such as in the
case of Newtonian liquids. Granular materials form
FFA, the cause of the linear increase of flow rates
measured in this work. The increase has not been
observed before, perhaps because it is obscured by the
Janssen effect. Our result shows that neither the
minimum energy theory nor the hour glass theory is
fully satisfactory, and the common view that the FFA
is able to resist all loads on it is at most an
approximation. As dissipative mechanisms are probably
responsible for the behavior of dynamic arches,
analyzing heat production should help to solve the
long-term mystery of GOF. Moreover, we observed that
pushing forces favorite a jamming transition, an
effect also operative in the slight increase of the
empty annulus size $D_{a}$ with the force.

\end{document}